\documentclass[preprint,amsmath,amssymb,aps,prd,showpacs]{revtex4-1}

\usepackage{epsfig}
\usepackage{dcolumn}
\usepackage{bm}
\usepackage{tikz}
\usepackage{graphicx, graphics, color}
\usepackage{dcolumn}
\usepackage{bm}
\usepackage{hyperref}
\usepackage[mathlines]{lineno}
\usepackage{float}
\usepackage{subfig}

\newcommand{\be}{\begin{equation}}
\newcommand{\ee}{\end{equation}}
\newcommand{\ben}{\begin{eqnarray}}
\newcommand{\een}{\end{eqnarray}}

\newcommand{\la}{{\lambda}}

\newcommand{\cO}{{\cal O}}

\newcommand{\cD}{{\cal D}}
\newcommand{\cM}{{\cal M}}
\newcommand{\cJ}{{\cal J}}

\newcommand{\p}{\partial}
\newcommand{\na}{\nabla}

\newcommand{\tF}{\tilde F}

\newcommand{\tB}{\tilde B}
\newcommand{\tA}{\tilde A}

\newcommand{\ep}{\epsilon}

\newcommand{\ga}{\gamma}

\begin{document}

\title{Uniqueness of dark matter magnetized static black hole spacetime}

\author{Marek Rogatko}
\affiliation{Institute of Physics \protect \\
Maria Curie-Sklodowska University \protect \\
20-031 Lublin, pl.~Marii Curie-Sklodowskiej 1, Poland \protect \\
rogat@kft.umcs.lublin.pl}

\date{\today}

\begin{abstract}
Uniqueness problem of static axially symmetric black hole in magnetic Universe filled with {\it dark matter} component has been considered.
The {\it dark matter} model comprises the additional $U(1)$-gauge field ({\it dark photon}) interacting with Maxwell one by the kinetic mixing term.
We show that all the solutions of Einstein-Maxwell {\it dark photon} gravity subject to the same boundary and regularity conditions
authorize the only static axially symmetric black hole solutions with non-vanishing time and azimuthal components of Maxwell and {\it hidden} sector gauge fields, 
say Schwarzschild-like black hole immersed in {\it dark matter} Melvin Universe.
\end{abstract}

\maketitle

\section{Introduction}
Elucidating the {\it dark matter} sector which comprises over 23 percent of the mass of observable Universe, and interacts
principally with ordinary ({\it visible sector}) through gravity, is one of the predominant pursuit in 
observational astrophysics, experimental high energy physics and theoretical attempts
of explaining its origin. Astrophysical observations reveal that non-baryonic cold {\it dark matter} comprises the dominant factor for the formation
of large-scale structures in the Universe, motion of galaxies and clusters of galaxies, as well as,
plays the crucial role in light bending coming from the outer space \cite{ber18}-\cite{bir16}.

On the other hand, our times are unforgettable for black hole physics, from the first Ligo gravitational wave detection to the 
Event Horizon Telescope (EHT) images of black hole shadow.

Studies of black holes in magnetic field comprise interesting problem on its own. Namely,
the effect of cosmological magnetic fields might lead to interesting astrophysical behaviors in the nearby of them. Secondly,
because of the fact that black hole magnetic solutions are not asymptotical flat ones, they also constitute interesting mathematical problem.
Moreover the problem of magnetic field in the nearby of black hole is interesting from the point of view of the recent measurements of it done by EHT team.

For the first time a regular static cylindrically symmetric solutions describing uniform magnetic field in general relativity were presented by Melvin in \cite{mel64, mel65}. 
Next, the problem of a rotating time-dependent magnetic universe was contemplated in \cite{gar94}, while the 
case of gravitational waves  and charged matter ones, traveling through magnetic universe
has elaborated in Refs. \cite{gar92, bin22}. The influence of the cosmological constant on the properties of Melvin universe has been analyzed in \cite{zof19, lin18}.

Black holes immersed in magnetic universe also attract much attention to. Namely, the magnetic Kerr and Kerr-Newman solutions were revealed in \cite{ern76},
while magnetized Kerr-Taub-NUT and Kerr-Newman-Taub-NUT solutions were elaborated in Refs. \cite{sia21, ghe21}.
On the other hand, the ultrarelativistic boosts of 
black holes in an external electromagnetic 
field was studied in \cite{ort04, ort18}, 
studies of ergoregions and thermodynamics of magnetized black hole were presented in \cite{gib13} (see also the earlier works 
connected with black holes in magnetized universe,
e.g., \cite{ear})).

Magnetized black hole solutions were also scrutinise in generalizations of Einstein theory of gravity, i.e.,
a Melvin universe with nontrivial dilaton and axion fields was founded \cite{tse95,gib88}, the dilaton C-metric in dilaton magnetic universe 
was presented \cite{dow94a}, whereas the case of a pair creation of the extremal black hole and Kaluza-Klein monopoles
was examined in \cite{dow94b}. On the other hand, Melvin-like solution with a Liouville type potential was given in Ref. \cite{rad04}, electrically charged dilaton black 
hole in magnetic field was analyzed in \cite{yaz13}, while
the generalization of the aforementioned problems to higher dimensional gravity was the subject of examination in Refs. \cite{ort05,yaz06}.

The interesting class of subjects was developed by using the Ernst's solution generation technique which enabled to scrutinize 
black hole solutions in magnetic universe in Einstein-Maxwell theory coupled 
conformally to a scalar field \cite{ast13}, axisymmetric stationary black holes with cosmological constant \cite{ast12},  
C-metric with conformally coupled scalar field in magnetic universe \cite{ast13b}, as well as,
regular solution describing a couple of charged spinning black hole in an external electromagnetic field \cite{ast14}.
Moreover, the solution unifying both magnetic Bertotti-Robinson and Melvin solution, as a single axisymmetric line element  was revealed \cite{maz13}.

Recent studies revealed that Melvin type solution could be found in gravity theories minimally coupled to any nonlinear electromagnetic theory, including 
Born-Infeld electrodynamics \cite{gib01}.

Furthermore gravitational collapse, physics of black holes, as well as, the uniqueness theorem for black holes, being the mathematical formulation of Wheeler's {\it black hole
no-hair} conjecture, focus much attention. Problem of classification of domains of outer communication of suitably regular black hole spacetimes 
in Einstein gravity has been widely elaborated in \cite{book}.

Higher dimensional generalization of gravity theory motivated by the contemporary unifications schemes 
such as M/string theories the classification of higher dimensional 
charged black holes both with non-degenerate and degenerate component 
of the event horizon has been exploited in \cite{nd}, while the nontrivial case 
of $n$-dimensional rotating black objects (black holes, black rings or black lenses) uniqueness theorem have been revealed
in \cite{nrot}. 

The quest for a consistent quantum gravity theory triggered interests 
in mathematical aspects of black holes in the low-energy limit of the string theories
and supergravity \cite{sugra}, and various modifications of Einstein gravity like
Gauss-Bonnet extension  \cite{shi13a, rog14}, Chern-Simons modified gravity \cite{shi13,rog13}, while the classification of static black holes in 
Einstein phantom-dilaton Maxwell--anti-Maxwell gravity systems has been given in \cite{rog22}.

On the other hand, the uniqueness theorem for black hole in magnetic Universe, in Ernst-Maxwell theory was elaborated in\cite{his81}, whereas
the magnetic Einstein-Maxwell dilaton gravity case was treated in \cite{rog16}.

Motivated by the above key problems, i.e., dark matter, black hole classifications and the influence of the magnetic field on the {\it no-hair theorem},
the main aim of our paper is to study the uniqueness of static black hole solutions in magnetic Universe. The additional point in our researches will be bounded with the influence of {\it dark sector}
on these objects. We shall pay attention to the so-called {\it dark photon} model, where the ordinary Maxwell gauge
field is supplemented by an auxiliary $U(1)$-gauge field, which interacts with Maxwell one by the {\it kinetic mixing} term. 

The organization of our paper is as follows. In Sec. I we describe the main assumptions leading to the {\it dark photon} theory and derive equations govern by {\it dark matter}.
Then, one rewrites the Einstein-Maxwell {\it dark photon} relations in the form of complex equations fulfilled by redefined gauge field strengths.
We also pay attention to the derivation of {\it dark matter} magnetic Melvin solution, which will be needed fot the uniqueness theorem for static
magnetized black holes with {\it dark matter} sector.
Sec. III is devoted to the boundary boundary conditions of the aforementioned equations of motion.
In Sec. IV we achieve the uniqueness of the static magnetized Schwarzschild-like black hole solution in Melvin Universe, say {\it dark} Melvin Universe Schwarzschild black hole.
Sec.V concludes our investigations.

\section{Equations of motion}
 The idea that {\it dark photon} can be a candidate for {\it dark matter} has been widely exploit on various backgrounds, both theoretically \cite{hol86}-\cite{abe08a} and experimentally \cite{til15}-\cite{an23}.
Additionally
 the model in question possesses some possible astrophysical confirmations \cite{jea03}-\cite{bod15}. We have cited only some illustrative examples due to the vast amount of work authorizing 
 this blossoming field of researches.
 
 To begin with let us consider the 
 Einstein-Maxwell {\it dark matter} gravity, where the {\it dark sector} will be described by the additional $U(1)$-gauge field ({\it dark photon}) coupled to the ordinary
 Maxwell one by the so-called {\it kinetic mixing} term, describing interactions of both gauge fields. The action related to Einstein-Maxwell {\it dark photon} gravity
 is provided by
 \be
S_{EM-dark~ photon} = \int \sqrt{-g}~ d^4x  \Big( R
- F_{\mu \nu} F^{\mu \nu} - B_{\mu \nu} B^{\mu \nu} - {\alpha}F_{\mu \nu} B^{\mu \nu}
\Big),
\label{act}
\ee  
where 
$\alpha$ is taken as a coupling constant between Maxwell and {\it dark matter} field strength tensors.

Introducing the redefined gauge fields $\tA_\mu$ and $\tB_\mu$, in the forms as follows:
\ben \label{ta}
\tA_\mu &=& \frac{\sqrt{2 -\alpha}}{2} \Big( A_\mu - B_\mu \Big),\\ \label{tb}
\tB_\mu &=& \frac{\sqrt{2 + \alpha}}{2} \Big( A_\mu + B_\mu \Big),
\een
one can get rid of the {\it kinetic mixing } term. Namely one arrives at
\be
 F_{\mu \nu} F^{\mu \nu} +
B_{\mu \nu} B^{\mu \nu} +  \alpha F_{\mu \nu} B^{\mu \nu}
\Longrightarrow
 \tF_{\mu \nu} \tF^{\mu \nu} +
\tB_{\mu \nu} \tB^{\mu \nu},
\ee
where we have denoted $\tF_{\mu \nu} = 2 \p_{[\mu }\tA_{\nu ]}$ and respectively $\tB_{\mu \nu} = 2 \p_{[\mu }\tB_{\nu ]}$. 

The rewritten action (\ref{act}) is given by
\be
S_{EM-dark~ photon}  = \int \sqrt{-g}~ d^4x  \Big( R
- \tF_{\mu \nu} \tF^{\mu \nu} - \tB_{\mu \nu} \tB^{\mu \nu}
\Big).
\label{vdc}
\ee
Variation of the action (\ref{vdc}) with respect to $g_{\mu\nu},~\tA_\mu$ and $\tB_\mu$ reveals the following equations of motion for Einstein-Maxwell {\it dark matter} gravity:
\ben
R_{\mu \nu} &=&
2 \tF_{\mu \rho} \tF_{\nu}{}{}^{\rho } - \frac{1}{2}g_{\mu \nu}\tF^{2} 
+ 2 \tB_{\mu \rho} \tB_{\nu}{}{}^{\rho } - \frac{1}{2}g_{\mu \nu}\tB^{2},\\
\na_{\mu} \tF^{\mu \nu } &=& 0, \qquad \na_{\mu} \tB^{\mu \nu } = 0.
\een
In what follows we shall consider the static axially symmetric background, due to the physical meaning of Melvin spacetime with magnetic field. In our case 
the magnetic field will be originated both from {\it visible} and {\it hidden} sector components.

The static axially symmetric line element under inspection yields
\be
ds^2 = - e^{2 \psi} dt^2 + e^{-2 \psi} \Big[ e^{2 \ga} \Big( dr^2 + dz^2 \Big) + r^2 d\phi^2 \Big],
\label{gij}
\ee
where we assume that the functions $\psi,~\ga$ depend on $r$ and $z$ coordinates.
On the other hand, the symmetry of the elaborated problem enforces that one supposes existence of time and azimuthal components of the $U(1)$-gauge
 fields. Consequently they yield
 \be
\tA_\mu dx^\mu = \tA_{t} dt + \tA_{\phi} d\phi, \qquad  \tB_\mu dx^\mu = \tB_{t} dt + \tB_{\phi} d\phi.
\ee
As the metric components, the gauge fields depend only on
$r$ and $ z$ coordinates.

The equations of motion for the considered {\it dark matter} Melvin axisymmetric spacetime are provided by
\be
\na^2 \psi - e^{-2\psi} 
\Big(
\tA_{t, r}^{2} + \tA_{t, z}^{2}  + \tB_{t, r}^{2} + \tB_{t, z}^{2} \Big) 
- \frac{e^{2\psi }}{r^2}
\Big( \tA^{2}_{\phi, r} + \tA^{ 2}_{\phi, z} + \tB^{2}_{\phi, r} + \tB^{ 2}_{\phi, z} \Big) = 0,
\label{emd1}
\ee
\be
\na_{r} \Big( r e^{- 2 \psi} \tA_{t, r} \Big) +
\na_{z} \Big( r e^{- 2 \psi}\tA_{t, z} \Big) = 0,
\label{emd2}
\ee
\be
\na_{r} \Big( r e^{- 2 \psi} \tB_{t, r} \Big) +
\na_{z} \Big( r e^{- 2 \psi} \tB_{t, z} \Big) = 0,
\label{emd2b}
\ee
\be
\na_{r} \Big(  \frac{e^{ 2 \psi }}{ r} \tA_{\phi, r} \Big) +
\na_{z} \Big(  \frac{e^{ 2 \psi}}{r} \tA_{\phi, z} \Big) = 0,
\label{emd3}
\ee
\be
\na_{r} \Big(  \frac{e^{ 2 \psi }}{ r} \tB_{\phi, r} \Big) +
\na_{z} \Big(  \frac{e^{ 2 \psi}}{r} \tB_{\phi, z} \Big) = 0,
\label{emd3b}
\ee
\be 
{\ga_{, z} \over r} - 2 \psi_{, r} \psi_{, z} =
- 2 e^{-2\psi} \tA_{t, r} \tA_{ t, z} + 
\frac{2}{ r^2} e^{2\psi} 
\tA_{ \phi, r} \tA_{\phi, z}  
- 2 e^{-2\psi} \tB_{t, r} \tB_{ t, z} + 
\frac{2}{ r^2} e^{2\psi} 
\tB_{ \phi, r} \tB_{\phi, z},
\label{emd4}
\ee
\be
e^{-2\psi } \Big(
\tA^{ 2}_{t, r} - \tA^{2}_{ t, z } + \tB^{ 2}_{t, r} - \tB^{2}_{ t, z }  \Big)
+
\frac{1}{r} e^{2\psi} 
\Big( \tA^{ 2}_{\phi, r} - \tA^{ 2}_{\phi, z} + \tB^{ 2}_{\phi, r} - \tB^{ 2}_{\phi, z} \Big) =
\psi_{, r}^2 - \psi_{, z}^2 - \frac{\ga_{, r}}{ r}.
\label{emd5}
\ee
The equations (\ref{emd3})-(\ref{emd3b}) can be regarded as integrability conditions for magnetic scalar potentials $\tA_3$ and $\tB_3$. Thus, one obtains
\be
\vec e_\phi \times \vec \na \tA_3 = \frac{e^{ 2 \psi }}{ r} \vec \na \tA_\phi, \qquad
\vec e_\phi \times \vec \na \tB_3 = \frac{e^{ 2 \psi }}{ r} \vec \na \tB_\phi,
\label{integr}
\ee
where for the brevity of the notation we set $\vec \na g = (\p_ r g,~\p_z g)$, while $\vec e_\phi $ is one of the orthonormal triad in the coordinate system $(r,~z,~\phi)$.
It can be noticed that he relations (\ref{integr}) lead to the conditions
\be
\p_r \p_z \tA_3 = \p_z \p_r \tA_3, \qquad \p_r \p_z \tB_3 = \p_z \p_r \tB_3.
\ee
Further, the expressions for $\vec e_\phi \times \vec \na \tA_\phi$ and $  \vec e_\phi \times \vec \na \tB_\phi$, have been used
and the complex potentials in the forms as, are defined
\be
\Phi_{(\tF)} = \tA_t + i~\tA_3, \qquad \Phi_{(\tB)} = \tB_t + i~\tB_3.
\label{poten}
\ee
In what follows, for the brevity of the notation, we rewrite potentials given by the expressions (\ref{poten}) in the forms given by
\ben \label{poo1}
\Phi_{(\tF)} &=& E_{(\tF)} + i ~B_{(\tF)},\\ \label{poo2}
\Phi_{(\tB)} &=& E_{(\tB)} + i ~B_{(\tB)}.
\een
In view of the definitions (\ref{poo1}) and (\ref{poo2}), Maxwell equations yield
\ben
\p_r \Big( r e^{-2 \psi} \p_r \Phi_{(\tF)} \Big) &+& \p_z \Big( r e^{-2 \psi} \p_z \Phi_{(\tF)} \Big) = 0,\\
\p_r \Big( r e^{-2 \psi} \p_r \Phi_{(\tB)} \Big) &+& \p_z \Big( r e^{-2 \psi} \p_z \Phi_{(\tB)} \Big) = 0,
\een
In the next step we define complex functions bounded with each of the gauge fields
\be
\ep_{(i)} = Z - \mid \Phi_{(i)} \mid^2 + i~Y_{(i)},
\label{poo3}
\ee
where $ Z = e^{2 \psi} $ and $i = \tF,~\tB$ and one introduces new potentials $k_{(i)}$, provided by
\be
\vec \na Y_{(i)} = - 2~Im \Big( \Phi^{\ast}_{(i)} \vec \na \Phi_{(i)} \Big).
\ee
Consequently the Einstein-Maxwell {\it dark matter} system of relations can be rewritten in a couple of complex equations,
fulfilled by each of the gauge field $\tF_{\mu \nu}$ and $\tB_{\mu \nu}$, which implies the following:
\ben \label{ern1}
\sum\limits_{i = \tF,\tB} \Big( Re~ \ep_{(i)}  +  \mid \Phi_{(i)} \mid^2 \Big) \na^2 \ep_{(i)} 
&=& \sum\limits_{i = \tF,\tB} 
\Big( \vec \na \ep_{(i)}  + 2 \Phi^{\ast}_{(i)} \vec \na \Phi_{(i)} \Big) \vec \na  \ep_{(i)},\\ \label{ern2}
\sum\limits_{i = \tF,\tB} \Big( Re ~\ep_{(i)}  +  \mid \Phi_{(i)} \mid^2 \Big) \na^2 \Phi_{(i)} 
&=& \sum\limits_{i = \tF,\tB} 
\Big( \vec \na \ep_{(i)}  + 2 \Phi^{\ast}_{(i)} \vec \na \Phi_{(i)} \Big) \vec \na  \Phi_{(i)}.
\een
The differential operators appearing in (\ref{ern1})-(\ref{ern2}) are defined as $\vec \na = (\p_r,~\p_z)$ and $\na^2 = (\p^2_r + \p^2_z + 1 /r ~\p_r)$ and constitute
flat gradient and Laplacian operators in cylindrical coordinates $(r,~z,~\phi)$, however in our considerations we restrict our attention to the functions depending on 
$(r,~z)$-coordinates.

As in the case of Ernst attitude to the Einstein-Maxwell system of differential equations, the real and imaginary parts of the above relations represent
the adequate equations of motion for the theory in question.
Moreover, it can be noticed that the effective action for stationary axisymmetric Ernst potentials given by
\be
S(\ep_{(i)},~\Phi_{(i)}) = \int dr dz \sum\limits_{i = \tF,\tB}
\frac{\Big( \vec \na \ep_{(i)} + 2 \Phi^{\ast}_{(i)} \na \Phi_{(i)} \Big) 
\Big(\vec \na \ep_{(i)}^\ast + 2 \Phi_{(i)} \vec \na \Phi^{\ast}_{(i)} \Big) }
{\Big( \ep_{(i)} + \ep_{(i)}^\ast + 2 \Phi^{\ast}_{(i)} \Phi_{(i)} \Big)^2},
\ee
leads to the aforementioned system of relations.

\subsection{Melvin dark Universe black hole}
In the latter section we arrive at the equation of motion for Einstein-Maxwell {\it dark photon} system. In order to proceed
to the uniqueness prove of static magnetized black hole solution in the theory under considerations,
one should specify the boundary conditions at infinity. As in the case of Einstein-Maxwell static black hole solutions, they tend asymptotically to  Melvin magnetic Universe solution
\cite{his81, rog16}. In the present case the magnetized static black hole solution with {\it dark photon} sector ought to tend asymptotically to
{\it dark} Melvin Universe one. Thus, firstly in this section we scrutinize the magnetostatic axisymmetric solution of Einstein-Maxwell {\it dark matter} system, we shall call {\it dark Melvin universe}.

The solution will be describe a static magnetic fields stemming from both {\it visible} and {\it dark} sectors. As in ordinary Melvin solution in general relativity,
the magnetic fields will be given as bundle of magnetic flux lines being in magnetostatic equilibrium with gravity. The Killing vectors of the underlying spacetime
are bounded with time translation symmetry, spatial translation along the axis, rotational symmetry, as well as, boost along the axis.
 Moreover,
the fields under considerations will have zero electric components, i.e., $\tF_{\alpha \beta} n^{\beta} = 0$ and $\tB_{\alpha \beta} n^{\beta} = 0$.

Having in mind the Maxwell equations of motion for both gauge field, we define pseudo-potentials
\ben \label{pot1}
\Phi^{(\tF)}_{,z} &=& -\frac{e^{2\psi}}{r} ~\tA_{\phi, r}, \qquad \Phi^{(\tF)}_{,r} = \frac{e^{2\psi}}{r} ~\tA_{\phi, z},\\ \label{pot2}
\Phi^{(\tB)}_{,z} &=&- \frac{e^{2\psi}}{r} ~\tB_{\phi, r}, \qquad \Phi^{(\tB)}_{,r} =  \frac{e^{2\psi}}{r} ~\tB_{\phi, z},
\een
which enables us to rewrite the equations as follows:
\ben
\frac{\p^2 \psi}{\p z^2} &+& \frac{1}{r} \frac{\p \psi}{\p r} + \frac{\p^2 \psi}{\p r^2} =
e^{- 2 \psi} \Bigg[ \Big( \frac{\p \Phi^{(\tF)}}{\p z} \Big)^2 + \Big( \frac{\p \Phi^{(\tF)}}{\p r} \Big)^2 + \Big( \frac{\p \Phi^{(\tB)}}{\p z} \Big)^2
+ \Big( \frac{\p \Phi^{(\tF)}}{\p z} \Big)^2 \Bigg], \\
\frac{1}{r} \frac{\p \gamma}{\p r} &-& 2 \frac{\p \psi}{\p r} \frac{\p \psi}{\p z} =
- 2 e^{- 2 \psi} \Bigg( \frac{\p \Phi^{(\tF)}}{\p z}  \frac{\p \Phi^{(\tF)}}{\p r}  +  \frac{\p \Phi^{(\tB)}}{\p z}  \frac{\p \Phi^{(\tB)}}{\p r} \Bigg),\\
\frac{1}{r} \frac{\p \gamma}{\p r} &+& \Big( \frac{\p \psi}{\p z} \Big)^2 - \Big( \frac{\p \psi}{\p r} \Big)^2 =
e^{- 2 \psi} \Bigg[ \Big( \frac{\p \Phi^{(\tF)}}{\p z} \Big)^2 - \Big( \frac{\p \Phi^{(\tF)}}{\p r} \Big)^2
+ \Big( \frac{\p \Phi^{(\tB)}}{\p z} \Big)^2 - \Big( \frac{\p \Phi^{(\tF)}}{\p z} \Big)^2 \Bigg],
\een
In order to obtain a bundle of magnetic flux lines, one assumes
that magnetic fields stemming from both gauge fields are directed along $z$-axis and the metric functions depend only on radial coordinates, one obtains 
\ben
\Phi^{(\tF)} &=& B^{(\tF)}_0 z, \qquad \Phi^{(\tB)} = B^{(\tB)}_0 z, \\
\psi(r) &=& \ln \Big[ 1 + \frac{1}{4} \Big( B^{(\tF) 2}_0 + B^{(\tB) 2}_0 \Big) ~r^2 \Big],\\
\ga(r) &=& 2~\ln \Big[ 1 + \frac{1}{4} \Big( B^{(\tF) 2}_0 + B^{(\tB) 2}_0 \Big) ~r^2 \Big],
\een
where $B^{(\tF)}_0$ and $B^{(\tB)}_0$ are constant bounded with the strength of the adequate magnetic fields, pertaining to both {\it visible} and {\it dark } sectors.

From equations (\ref{pot1}) and (\ref{pot2}) one finds that
\ben
\tA_\phi &=&  \frac{2 B^{(\tF)}_0 }{\Big( B^{(\tF) 2}_0 + B^{(\tB)2}_0 \Big)}
\frac{1}{\Big[1 + \frac{1}{4} \Big( B^{(\tF) 2}_0 + B^{(\tB) 2}_0 \Big) ~r^2 \Big]},\\
\tB_\phi &=&  \frac{2B^{(\tB)}_0}{\Big( B^{(\tF) 2}_0 + B^{(\tB)2}_0 \Big)}
\frac{1}{
~ \Big[1 + \frac{1}{4} \Big( B^{(\tF) 2}_0 + B^{(\tB) 2}_0 \Big) ~r^2 \Big]}.
\een

It can be seen that the {\it dark photon} field influences the obtained potentials.
In order to envisage the {\it dark photon} impact let us use the relation (\ref{ta}) and (\ref{tb}) and rewrite $B^{(\tF)}_0$ and $B^{(\tB)}_0$ as follows:
\be
B^{(\tF)}_0 = \frac{\sqrt{2 - \alpha}}{2} \Big( B^{(F)}_0 - B^{(B)}_0 \Big),
\qquad
B^{(\tB)}_0 = \frac{\sqrt{2 + \alpha}}{2} \Big( B^{(F)}_0 + B^{(B)}_0 \Big),
\ee
where $B^{(F)}_0$ denotes constant Maxwell magnetic field and $B^{(B)}_0$ stands for constant {\it dark photon} magnetic component.
After some algebra we obtain
\ben
\tA_\phi &=& \frac{\sqrt{2 - \alpha}}{2} \Big( P^{(F)} - P^{(B)} \Big),\\
\tB_\phi &=& \frac{\sqrt{2 + \alpha}}{2} \Big( P^{(F)} + P^{(B)} \Big),
\een
where we set
\ben
 P^{(F)} &=& \frac{2 B^{(F)}_0}{\Big[1 + \frac{1}{4} \Big( B^{(F) 2}_0 + B^{(B) 2}_0 \Big) ~r^2 \Big]},\\
 P^{(B)} &=& \frac{2 B^{(B)}_0}{\Big[1 + \frac{1}{4} \Big( B^{(F) 2}_0 + B^{(B) 2}_0 \Big) ~r^2 \Big]}.
\een
On the other hand,
at large distances of $r$, the obtained metric reveals that $\psi(r) \simeq 2 \ln r$, as in ordinary Einstein-Maxwell Melvin case \cite{mel65}. However at infinity, {\it dark Melvin Universe}
solution approaches to a non-flat solution, which will constitute the crucial point in the boundary conditions and then in uniqueness theorem.

\section{Boundary conditions}
This section will be devoted to the relevant boundary conditions in the case
under consideration. 
In the present case the spacetime is asymptotically cylindrical, i.e., the static magnetized black hole solution will tend asymptotically to {\it dark Melvin Universe},
describing the bundle of magnetic flux lines. This fact constitutes the main difference between the studied case and the asymptotically flat one.

To begin with we introduce the two-dimensional manifold, equipped with the spheroidal coordinates provided by the relations
\be
r^2 = (\la^2 - c^2)~(1 - \mu^2), \qquad z = \la~\mu,
\label{coor}
\ee
where $\mu = \cos \theta$ is chosen in such a way that the black hole event horizon boundary is situated 
at a constant value of $\la = c$.
On the other hand, two rotation axis segments which distinguish the south and the north segments of the event horizon
are described by the respective limit $\mu = \pm 1$.
Just we obtain the line element in the form as
\be
dr^2 + dz^2 = (\la^2 - \mu^2~c^2)\bigg(
{d\la^2 \over \la^2 - c^2} + {d\mu^2 \over 1 - \mu^2} \bigg).
\ee

We choose the domain of outer communication $<<\cD>>$ as a rectangle
\ben \label{dom}
\p \cD^{(1)} &=& \{ \mu = 1,~\la= c, \dots, R \},\\ \nonumber
\p \cD^{(2)} &=& \{ \la = c,~\mu = 1, \dots, -1 \},\\ \nonumber
\p \cD^{(3)} &=& \{ \mu = - 1,~\la= c, \dots, R \},\\ \nonumber
\p \cD^{(4)} &=& \{ \la  = R,~\mu= -1, \dots, 1 \}.
\een

The relevant boundary conditions may be cast as follows.
At infinity, we insist that
$Z,~\tA_\phi, ~\tA_t$ and $~\tB_\phi, ~\tB_t$
are well-behaved functions and the solution under inspection asymptotically tends
to the Melvin {\it dark matter} Universe line element, presented in the preceding section. Namely, they satisfy

\ben \label{inf1}
Z &=&  \Bigg[1 +\frac{1}{4} \Big( B^{(\tF) 2}_0 + B^{(\tB) 2}_0 \Big) ~r^2 \Bigg]^{2}~\Big( 1 + \cO(\la^{-1})\Big),\\ \label{inf2}
\tA_\phi &=& \frac{2 B^{(\tF)}_0}{\Big( B^{(\tF) 2}_0 + B^{(\tB)2}_0 \Big)}
 \Bigg[1 + \frac{1}{\frac{1}{4} \Big( B^{(\tF) 2}_0 + B^{(\tB) 2}_0 \Big)~ r^2} \Bigg]^{-1}~
\Big( 1 + \cO(\la^{-1})\Big),\\ \label{inf3}
\tA_t &=& \cO(\la^{-1}), \\ \label{inf4}
\tB_\phi &=& \frac{2 B^{(\tB)}_0}{\Big( B^{(\tF) 2}_0 + B^{(\tB)2}_0 \Big)}
\Bigg[1 + \frac{1}{\frac{1}{4} \Big( B^{(\tF) 2}_0 + B^{(\tB) 2}_0 \Big)~r^2} \Bigg]^{-1}~
\Big( 1 + \cO(\la^{-1})\Big),\\ \label{inf5}
\tB_t &=& \cO(\la^{-1}),
\een
where now $r$ stands for the asymptotical cylindrical coordinate given by
$r^2 \rightarrow \la^2~(1 - \mu^2)$. 
The difference in the boundary behaviors at infinity comprises the main distinction between the considered case and the asymptotically flat one. 
As was previously mentioned the solution should display asymptotically {\it dark} Melvin Universe one, in order to reveal the cylindrical nature of the elaborated spacetime.

On the black hole event horizon,
where $\la \rightarrow c$, the quantities in question should behave regularly (see, e.g., \cite{book}), i.e.,
they yield the following relations:
\ben \label{hor}
Z &=& \cO(1 - \mu^2), \qquad \frac{1}{ Z}~\p_\mu~Z = - \frac{2 \mu}{ 1 - \mu^2} + \cO(1),\\
\p_\la~\tA_\phi &=& \cO(1 - \mu^2), \qquad \p_\mu~\tA_\phi = \cO(1), \\
\p_\la~\tA_t &=& \cO(1), \qquad  \p_\mu~\tA_t = \cO(1), \\
\p_\la~\tA_\phi &=& \cO(1 - \mu^2), \qquad \p_\mu~\tA_\phi = \cO(1), \\
\p_\la~\tA_t &=& \cO(1), \qquad  \p_\mu~\tA_t = \cO(1), \\
\p_\la~\tB_\phi &=& \cO(1 - \mu^2), \qquad \p_\mu~\tB_\phi = \cO(1), \\
\p_\la~\tB_t &=& \cO(1), \qquad  \p_\mu~\tB_t = \cO(1), \\
\p_\la~\tB_\phi &=& \cO(1 - \mu^2), \qquad \p_\mu~\tB_\phi = \cO(1), \\
\p_\la~\tB_t &=& \cO(1), \qquad  \p_\mu~\tB_t = \cO(1).
\een
On the other hand, in the vicinity of the symmetry axis, where  $\mu \rightarrow 1$ (north polar segment) and $\mu \rightarrow -1$ (south polar segment),
one requires that $\tA_\phi, \tA_t,~\tB_\phi,~\tB_t, ~Z$ should be regular functions of $\la$ and $\mu$, such that
\ben \label{ax}
Z &=& \cO(1), \qquad \frac{1}{Z} = \cO(1),\\
\tA_\phi &=& \cO(1), \qquad \p_\la~\tA_\phi = \cO(1), \\
\tA_t &=& \cO(1), \qquad \p_\la \tA_t = \cO(1),\\
\tB_\phi &=& \cO(1), \qquad \p_\la~\tB_\phi = \cO(1), \\
\tB_t &=& \cO(1), \qquad \p_\la \tB_t = \cO(1).
\een

\section{Uniqueness of Solutions}
To commence with, we recall that the Ernst equations of the type described by relations (\ref{ern1}) and (\ref{ern2}), can be cast in the matrix type system of equations
\be
\p_{r} \Big[ P^{-1} \p_{r} P \Big]
+ \p_{z} \Big [ P^{-1} \p_{z} P\Big ] = 0,
\label{mat}
\ee
where $P$ are $3\times 3$ Hermitian matrices with unit determinants, while
including parts bounded with the adequate gauge fields, described by the field strengths $\tF_{\mu \nu},~ \tB_{\mu \nu}$.
For the first time this problem was investigated in \cite{gur82}.

Moreover if one considers any constant invertible matrix $A$, the matrix built in the form as $A P A^{-1}$ constitute the solution of (\ref{mat}).
The different forms of these matrices enable to construct all the transformations referred to the Ernst's system of partial differential equations.

To proceed further let us examine a domain of outer communication $<<\cD>>$ of the two-dimensional manifold $\cM$, with boundary $\p \cD$.
Suppose next, that the matrix $P$ components are enough differentiable in the domain of outer communication in question.
Let us inspect the two different matrix solutions of (\ref{mat}), i.e., $P_{1}$ and $P_{ 2}$, subject to the same boundary and differentiability
conditions.

The difference between the aforementioned relations fulfils the equation as follows:
\be
\na \Big(  P_{ 1}^{-1} \Big( \na Q \Big)
P_{ 2} \Big) = 0, 
\label{diff}
\ee
where we have denoted by $Q = P_{1} P_{2}^{-1}$. In the next step one can multiply the equation (\ref{diff}) by
$Q^{\dagger}$ and taking the trace of the result. One arrives at the following:
\be
\na^2 q = Tr \Bigg[
\Big( \na Q^{\dagger} \Big) P_{ 1}^{-1}
\Big ( \na Q \Big) P_{ 2} \Bigg].
\label{tr}
\ee
In the above relation we set $q  = Tr Q $. Further the hermicity and positive definiteness of $P$ allow to postulate the form
of it given by $P = M  ~M^{\dagger}$, which leads to the relation
\be
\na^2 q = Tr \Big( \cJ^{\dagger} \cJ \Big),
\label{jj}
\ee
where $ \cJ= M_1^{-1} (\na Q ) M_{ 2}$.

Defining homographic change of the variables, for the previously defined quantities connected with both gauge fields, provided by
\be
\ep_{(i)} = \frac{ \xi_{(i)} -1}{\xi_{(i)} + 1}, \qquad \psi_{(i)} = \frac{\eta_{(i)}}{\xi_{(i)} +1},
\ee
enables us to find that the $P$ matrix implies
\be
P_{\alpha \beta } = \eta_{\alpha \beta} - \frac{2 \xi_\alpha~ {\bar \xi}_\beta}{<\xi_\delta~ {\bar \xi}^\delta>},
\ee
where we define the scalar product in the form as
\be
<\xi_\delta~ {\bar \xi}^\delta> = -1 + \sum \limits_\ga \xi_\ga~ {\bar \xi}^\ga, \qquad \ga= 1, \dots, q.
\ee
To proceed to the uniqueness of the considered solution, one has to calculate $q_{(i)} = Tr(P_1 P^{-1}_2)$, having in mind the adequate Ernst's potentials $\ep_{(i) (1)}$ and 
$\ep_{(i) (2)}$ and their ingredients given by the equations (\ref{poo1})-(\ref{poo2}) and (\ref{poo3}), like $E_{(i)1},~B_{(i)1},~Y_{(i)1},~Z_{(i)1}$
and $E_{(i)2},~B_{(i)2},~Y_{(i)2},~Z_{(i)2}$,
 where $i =\tF,~\tB$.
Consequently, after some algebra, one achieves the relation provided by
\ben \label{qq} \nonumber
q &=&  P_{\alpha \beta (1)} P^{\alpha \beta}_{(2)} =3 + \frac{1}{Z_1 Z_2} \sum \limits_{i = \tF, \tB} \Bigg[
(Z_1 - Z_2)^2 + \frac{1}{4} \Big[ (E_{(i)1} -E_{(i)2})^2 + (B_{(i)1} - B_{(i)2})^2 \Big]^2 \\ \nonumber
&+&
(Z_1 + Z_2) \Big[ (E_{(i)1} -E_{(i)2})^2 + (B_{(i)1} - B_{(i)2})^2  \Big]  \\ 
&+& \Big[ \Big( B_{(i)1} E^{(i)}_2 - B_{(i)2} E^{(i)}_1 \Big) + \frac{1}{2} (Y_{(i) 1} - Y_{(i)2} ) \Big]^2 
\Bigg].
\een

On the other hand, the use of the Stoke's theorem authorizes the integration of the relation given by (\ref{jj}) over the chosen
domain of outer communication $<<\cD>>$, described by the relation (\ref{dom}), reveals
\ben \label{bou} \nonumber
\int_{\p <<\cD>>} \na_m q~dS^m &=&   
\int_{\p <<\cD>>}~d\la~\sqrt{{h_{\la \la} \over h_{\mu \mu}}}~\p_\mu q \mid_{\mu = const}
+ \int_{\p <<\cD>>}~d\mu~\sqrt{{h_{\mu \mu} \over h_{\la \la}}}~\p_\la q \mid_{\la = const}
\\
&=&
\int_{\infty}^{c} ~d \la \sqrt{{h_{\la \la} \over h_{\mu \mu}}}~\p_\mu q \mid_{\mu = -1} + \int_{c}^{\infty} ~d \la \sqrt{{h_{\la \la} \over h_{\mu \mu}}}~\p_\mu q \mid_{\mu = 1} \\ \nonumber
&+&
\int_{1}^{-1} ~d \mu \sqrt{{h_{\mu \mu} \over h_{\la \la}}}~\p_\la q \mid_{\la = c} + \int_{-1}^{1} ~d \mu \sqrt{{h_{\mu \mu} \over h_{\la \la}}}~\p_\la q \mid_{\la \rightarrow \infty} \\ \nonumber
&=&
\int_{<<\cD>>} Tr \Big( \cJ^{\dagger} \cJ \Big)~dV.
\een
To proceed further, we ought to indicate the behavior of the left-hand side of the equation (\ref{bou}), taking into account the integrals over each
part of the domain of outer communication $ <<\cD^{(i)}>>$, where $i = 1,\dots,4$, taken as the rectangle in the two-dimensional space $(\mu,~\la)$ and described by
the relations (\ref{dom}).

On the black hole event horizon, $\p \cD^{(2)}$, 
the functions 
are well behaved, with the asymptotic given by $\cO(1)$. For the $r$-coordinate given by the equation (\ref{coor})
we have that $r \simeq \cO( \sqrt{\la -c})$ as $\la \rightarrow c$. Moreover, the square root $\sqrt{\frac{h_{\mu \mu}}{h_{\la \la}}} \simeq \cO( \sqrt{\la -c})$. It all leads to the conclusion that
$\na_m q$ vanishes on the black hole event horizon.

On the other hand, on the symmetry axis, $\p \cD^{(1)}$ and $\p \cD^{(3)}$,
when $\mu \pm 1$, all the considered quantities are of order $\cO(1)$. 
The $r$-coordinate tends to $\cO( \sqrt{1 - \mu})$, as $\mu \rightarrow 1$. When $\mu \rightarrow - 1$,
one obtains that $r \simeq \cO( \sqrt{1 + \mu})$. As far as the behavior of the square roots is concerned they are provided by
$\sqrt{\frac{h_{\la \la}}{h_{\mu \mu}}} \simeq \cO( \sqrt{1 + \mu})$, when $\mu \rightarrow -1$ and
$\sqrt{\frac{h_{\la \la}}{h_{\mu \mu}}} \simeq \cO( \sqrt{1 - \mu})$, when $\mu \rightarrow 1$. Having in mind the 
relations (\ref{qq})) enables us to conclude that $\na_m q = 0$, for $\mu \pm 1$.

It remains to take into account the contribution for the integration along  $\p \cD^{(4)}$, when
$\la = R \rightarrow \infty$.
  We remark that the main difference between the case under consideration and asymptotically flat one, is in the boundary conditions
 at infinity, where we insist that all functions in question are well-behaved and have asymptotic behaviors given by the equations (\ref{inf1})-(\ref{inf5}).
  The square root in the considered limit tends to $\sqrt{\frac{h_{\mu \mu}}{h_{\la \la}}} \simeq \cO(\la)$.

Let us analyze $q$ given by the equation (\ref{qq}), term by term, in the considered limit of $\la$. They will have the same behavior for each gauge field, described by $\tF_{\mu \nu}$
and $\tB_{\mu \nu}$.
The first one is equal to the constant value, while the second one is given by
\be
\frac{\Big( Z_1 - Z_2 \Big)^2}{Z_1 Z_2} \simeq
\frac{\Delta B_{(1)}^4 - \Delta B_{(2)}^4}{\Delta B_{(1)}^4~\Delta B_{(2)}^4} \Big( 1 + \cO\Big( \frac{1}{\la} \Big) \Big),
\ee
where we have denoted
\be
\Delta B_{(i)}^2 = B_{0 (i)}^{(\tF,\tB) 2} + B_{0 (i)}^{(\tB,\tB) 2},
\ee
$i = 1,~2$.
 For  the third term in equation (\ref{qq}) one obtains the relation
 \be
\frac{\frac{1}{4} \Big[ (E_{(\tF,\tB)1} -E_{(\tF,\tB)2})^2 + (B_{(\tF,\tB)1} - B_{(\tF,\tB)2})^2 \Big]^2 }{Z_1 Z_2} \simeq
\frac{\mu^4 ~\Big( B_{0(1)}^{(\tF, \tB )} - B_{0(1)}^{(\tF, \tB )} \Big)^2}{\Delta B_{(1)}^4~\Delta B_{(1)}^4~\Big(1 - \mu^2 \Big)^2} ~\cO\Big( \frac{1}{\la^6} \Big),
\ee
while the fourth one implies
\ben
\frac{(Z_1 + Z_2)}{Z_1 Z_2}~ \Big[ (E_{(\tF,\tB)1} -E_{(\tF,\tB)2})^2 + (B_{(\tF,\tB)1} - B_{(\tF,\tB)2})^2  \Big]   \\ \nonumber
\simeq \frac{
\mu^4~\Big( \Delta B_{(1)}^4 +  \Delta B_{(2)}^4 \Big)~\Big( B_{0(1)}^{(\tF,\tB)} - B_{0 (2)}^{(\tF,\tB)} \Big)^2}
{\Delta B_{(1)}^4 ~\Delta B_{(2)}^4 } \cO\Big( \frac{1}{\la^2}\Big).
\een
For the last one, the fifth term, it can be revealed that
\ben
\frac{
  \Big[ \Big( B_{(\tF,\tB)1} E^{(\tF,\tB)}_2 - B_{(\tF,\tB)2} E^{(\tF,\tB)}_1 \Big) + \frac{1}{2} (Y_{(\tF,\tB) 1} - Y_{(\tF,\tB)2} ) \Big]^2 }{Z_1 Z_2}\\ \nonumber
  \simeq
\frac{
\Big( B_{0(1)}^{(\tF,\tB)} - B_{0 (2)}^{(\tF,\tB)} \Big)^2~\Big( 1 - 2 \mu^2 \Big)}
{\Delta B_{(1)}^4 ~\Delta B_{(2)}^4 ~\Big(1 - \mu^2 + \mu \Big)^2~\Big( 1 - \mu^2 \Big)^2} \cO \Big( \frac{1}{\la^6}\Big).
\een
Consequently the function $q$ displays the following way of acting:
\be
q \mid_{\la \rightarrow \infty} \simeq \cO(1) + \cO\Big( \frac{1}{\la^6} \Big) + \cO\Big( \frac{1}{\la^2} \Big) + \cO\Big( \frac{1}{\la^6} \Big) \simeq \cO( 1).
\ee
In view of the above relations, one has that $q$ tends to a constant value, as $\la \rightarrow \infty$.
All the aforementioned arguments lead to the conclusion that
\be
\int_{<<\cD>>} Tr \bigg( \cJ_{(i)}^{\dagger} \cJ_{(i)} \bigg) = 0.
\label{jj1}
\ee    
This relation implies that $P_{(i) 1} = P_{(i) 2}$, at all points belonging to the domain of outer communication, being a two-dimensional manifold
$\cM$ with coordinates $(r,~z)$.

Thus if we consider two black hole solutions of Einstein-Maxwell {\it dark photon} gravity characterized respectively by
$(Z_{(1)},\tA_{t (1)},\tA_{\phi (1)}, \tB_{t (1)},\tB_{\phi (1)})$ and 
$(Z_{(2)},\tA_{t (2)},\tA_{\phi (2)}, \tB_{t (2)},\tB_{\phi (2)})$, being subject to the same boundary and regularity conditions, they are identical.

In summary, the consequences of our research can be summarized as follows:\\
\noindent
{\bf Theorem}:\\
Let $ <<\cD>>$ be a domain of outer communication constituting a region of two-dimensional manifold with coordinates $(r,~z)$ defined by (\ref{coor}), having
a boundary $ <<\p \cD>>$.  Suppose, that  $P_{(i)}$ are Hermitian positive, three-dimensional matrices, with unit determinants.
On the boundary of the domain $ <<\p \cD>>$, matrices $P_{(1)}$ and $ P_{(2) }$ being the solution of the equation
$$\p_{r} \Big[ P^{-1} \p_{r} P \Big]
+ \p_{z} \Big [ P^{-1} \p_{z} P\Big ] = 0,$$
 satisfy the relation $\na_m q =0$. Then, $P_{(1)} = P_{(2)}$ in all domain of outer communication $ <<\cD>>$, implying that for at least one point $d \in <<\cD>>$,
 one has that 
 $ P_{(1) } (d) = P_{(2)} (d). $\\

In other words, all the solutions of Einstein-Maxwell {\it dark photon} gravity subject to the same boundary and regularity conditions,
say {\it dark Melvin Universe} Schwarzschild type black hole, comprise the only static, axisymmetric symmetric black hole solution,
possessing a regular event horizon, with non-vanishing
$\tA_{t },~\tA_{\phi },~ \tB_{t},~\tB_{\phi }$ components of Maxwell {\it visible}  and {\it hidden sectors} gauge fields.

\section{Conclusions}
Our paper is devoted to the uniqueness  problem of static axially symmetric black hole spacetime
in Einstein-Maxwell {\it dark photon} gravity.

{\it Dark photon} model comprises a new Abelian gauge field coupled to the ordinary Maxwell one, by the kinetic mixing term. The model
in question is subject of extensive theoretical and experimental studies.

The equations of motion for {\it dark photon} Einstein-Maxwell gravity can be rewritten in the form of Ernst-like system of complex relations,
which can be rearranged in the form of matrix equations. In the studies,
the domain of outer communication $ <<\cD>>$ was chosen as a two-dimensional manifold with coordinates $(r,~z)$.
It has been revealed that the two matrix solutions of the equations of motion, subject to the same boundary and regularity conditions,
are equal in the considered domain of outer communication. 

One may conclude that Schwarzschild
Melvin-like solution with {\it dark photon} (representing model of {\it dark matter} sector), is the only axisymmetric static 
black hole in Einstein-Maxwell  {\it dark photon} gravity with non-zero components
of {\it visible} Maxwell and {\it hidden sector} $U(1)$-gauge components provided by
$\tA_{t },~\tA_{\phi },~ \tB_{t},~\tB_{\phi }$, being Schwarzschild type black hole immersed in magnetic Melvin Universe, filled with {\it dark matter}.



\acknowledgements
M.R. was partially supported by Grant No. 2022/45/B/ST2/00013 of the National Science Center, Poland.



\end{document}